# Surface magnetism of gallium arsenide nanofilms

Huan Lu, Jin Yu, Wanlin Guo*

*State Key Laboratory of Mechanics and Control of Mechanical Structures and Key Laboratory for Intelligent Nano Materials and Devices of MOE, Institute of Nano Science, Nanjing University of Aeronautics and Astronautics, Nanjing, 210016, China*

**Abstract:** Gallium arsenide (GaAs) is the widest used second generation semiconductor with a direct band gap and increasingly used as nanofilms. However, the magnetic properties of GaAs nanofilms have never been studied. Here we find by comprehensive density functional theory calculations that GaAs nanofilms cleaved along the <111> and <100> directions become intrinsically metallic films with strong surface magnetism and magnetoelectric (ME) effect. The surface magnetism and electrical conductivity are realized via a combined effect of transferring charge induced by spontaneous electric-polarization through the film thickness and spin-polarized surface states. The surface magnetism of <111> nanofilms can be significantly and linearly tuned by vertically applied electric field, endowing the nanofilms unexpectedly high ME coefficients, which are tens of times higher than those of ferromagnetic metals and transition metal oxides.

---

*Corresponding author: Tel. +86 25 84891896; Fax. +86 25 84895827; email: wlguo@nuaa.edu.cn

# I. INTRODUCTION

Magnetism originating from surfaces and interfaces is always related to strong correlated systems with unoccupied *d* or *f* electrons, and has attracted great attention in device development. In the last century, this phenomenon has been intensively investigated in magnetic metals and transition metal oxides [1-5], and is usually attributed to the RKKY interaction [6-8] or spin-dependent exchange interaction [9-11]. In recent decades, the magnetic system has been extended to *s* and *p* hybrid electrons with the rise of low-dimensional materials [12-18]. In graphene, BN and other two-dimensional crystals, magnetism can be induced by defects, structure distortions [19-21] as well as edge states [12,22,23]. Recently, ferromagnetism on reconstructed Si<111> surfaces has been theoretically predicted, where the time-reversal symmetry is broken by the spontaneous surface reconstruction and magnetic instability [24]. Magnetic moments in metal oxides and perovskite materials caused by holes in oxygen p orbitals have also been widely predicted [25-30], and ferromagnetic ordering is obtained if the hole density is high enough [25,26,30-32]. A representation is the polar (0001) oriented surfaces of wurtzite ZnO, in which local spin polarization of O atoms induced at the surface is 3 times larger than in the bulk [31], and the surface ferromagnetism can been considerably tuned via doping Co [31] or hydrogen adsorption [33]. Oxygen has a high electronegativity, which makes the surface oxygen easily achieve the high density of states at the Fermi level. However, it is challenging for other atoms on the surface to achieve that. One possible strategy is doping carriers to increase the density, and it has been realized in monolayer GaSe [34]. Charge transfer is another effective way to induce magnetism [35-39], in $(LaNiO_3)_n/(LaMnO_3)_2$ superlattices, an interfacial magnetism is realized in $LaNiO_3$ with electrons transferring from the $LaMnO_3$ to the $LaNiO_3$ [39].

Usually, the surface magnetism can be tuned by vertically applied electric field, which is called surface ME effect [40]. Up to now, there are almost 100 compounds having been studied to reveal the ME effect [41-43]. The underlying mechanisms of the ME effect can be classified into two categories. In ferromagnetic metal films and graphene nanoribbons [44-46], the ME effect results from the electric field-induced spin imbalance and exchange interaction. As for multi-ferroelectric or ferroelectric-ferromagnetic multilayers [47-51], the ME effect is by virtue of the piezoelectric strain in the ferroelectric constituent of the heterostructure, which would change the magnetic properties of the ferromagnetic constituent [52-54]. As ME effects in these materials are confined to the interfaces or surfaces, the relationship between the induced magnetization and external electric field can be expressed as

$$\mu_0 \Delta \mathrm{M} = \alpha \mathrm{E}, \qquad (1)$$

Where $\mu_0$ denotes the magnetic permeability of vacuum and $\alpha$ denotes the surface or interface ME coefficient.

Besides changing magnetic properties, surface states can also distinguish nanofilms from their bulk materials in electrical conductivity. The electronic reconstruction at the surface or interface can give rise to a highly-conductive property [55-58], and sometimes it combines with magnetism [57,58]. An representative case in this area is cubic boron nitride (BN) <111> nanofilms [59]. In contrast to intrinsic electrical insulation of BN materials, the BN nanofilms become metallic because of the labile near-gap states originating from the surface.

As an important second generation semiconductors, sphalerite GaAs is non-magnetic with a direct band gap of 1.43 eV and more and more used as nanofilms [60]. The main efforts so far have focused on GaAs-based diluted magnetic semiconductors [61-65], which are considered as strong candidates for the room-temperature magnetic semiconductor [66]. Doping magnetic atoms to GaAs surfaces results in various magnetic properties as well [67-69]. A trend has been revealed that the magnetic anisotropy energy is a function of the cluster size for an individual Mn impurity positioned in the vicinity of the <110> GaAs surface [67]. Nonetheless, the intrinsic surface states of GaAs nanofilms have never been reported. In this study, we find that GaAs nanofilms cleaved along the <111> and <100> directions become intrinsically metallic films with strong surface magnetism and ME effect. With charge-transfer normal to the nanofilms, excess screening charge confined to a depth of a few atoms from the surface leads to the intrinsic metallicity of the whole structure. Due to exchange interactions, the screening charge is spin-dependent, exhibiting surface magnetism. Once these nanofilms are exposed to vertically applied electric field, the spin dependence of the screening electrons leads to a strong ME effect. Since the electric field hardly penetrates into mid-layers of GaAs nanofilms, the ME effect is limited to the surface as surface ME effect.

## II. CALCULATION METHOD

All calculations are carried out based on density-functional theory (DFT) in the Vienna *ab initio* simulation package (VASP) [70,71]. The studied systems are free-standing fcc GaAs<111> nanofilms and fcc GaAs<100> nanofilms with periodic boundary conditions. The Kohn-Sham equation was solved iteratively using a plane wave basis set with a cutoff energy of 500 eV to describe the valence electrons. The exchange correlation effects were incorporated in the spin-polarized generalized gradient approximation (GGA) using the Perdew–Burke–Ernzerhof (PBE) functional, and the electron-ion interactions were described by the projector augmented wave (PAW) method [72]. For the hexagonal unit cell, the Brillouin-Zone sampling was performed using a 15×15×1 MP grid for atomic structure relaxation calculations and a 30×30×1 MP grid for static calculations [73]. All of the atoms in the unit cell were fully relaxed until the force on each atom was less than 0.001 eV/Å. Electronic minimization was performed with a tolerance of $10^{-5}$ eV. The vacuum between two adjacent planes was larger than 15 Å to separate the interaction between periodic images. The uniform external electric field applied perpendicular to the nanofilm surface was introduced by planar dipole layer method as implemented in VASP [74]. The dipole correction [75] is applied to set the electric field in the vacuum region to zero.

## III. RESULTS AND DISSCUSIONS

The nanofilms are cleaved along the <111> and <100> directions of the cubic GaAs structure without passivation, the unit cells of which are standard rhombus with optimized lattice constant of a = b = 3.997 Å and quadrate with a = b = 4.093 Å. The thickness of nanofilms is defined as the cleaved monolayer (ML) number **n** indexed by subscript. The outmost surface with Ga atom is denoted as the Ga-surface, and the outmost surface with As atom is denoted as the As-surface. First of all, the stability of the nanofilms is confirmed by quantum *ab initio* molecular-dynamics calculations, see Supplemental Material Fig. S1. Actually, the energetic and dynamical stability of single-layer III-V materials has been confirmed by first principle calculations [75]. Optimized atomic structures of one unit <111> and <100> nanofilms are presented in Figs. 1(a) and 1(b), and the bond length analysis is in Figs. 1(c) and 1(d). The surface atoms of <111> nanofilms have little fluctuation, and the bond length of surface layers varies much more than that of the mid-layers for **n** from 7 to 9. It is worth mentioned that the bond length of <111> nanofilms is never fixed to the bulk value, which is about 2.50 Å in our simulations, even in the center of the slab as thick as 9 MLs. As presented in Fig. 1(a), those bonds almost perpendicular to the nanofilms are slightly shorter than the bonds of others because of the built-in electric field along the same direction, which will be described below. Further, when spin-polarization is taken into account, the bond length of the outmost Ga and the closest As atom change slightly, rendering a more undulating Ga-surface. The bond length of <100> nanofilms has the similar phenomenon, see Fig. 1(d). As being reported in many wurtzite nanostructures that it is hard for those nanostructures to form graphitic layers with a few atom layers when **n** is less than three [76,77], GaAs nanofilms show the same phenomenon as well. Therefore, we focus on the magnetic properties of those thicker nanofilms.

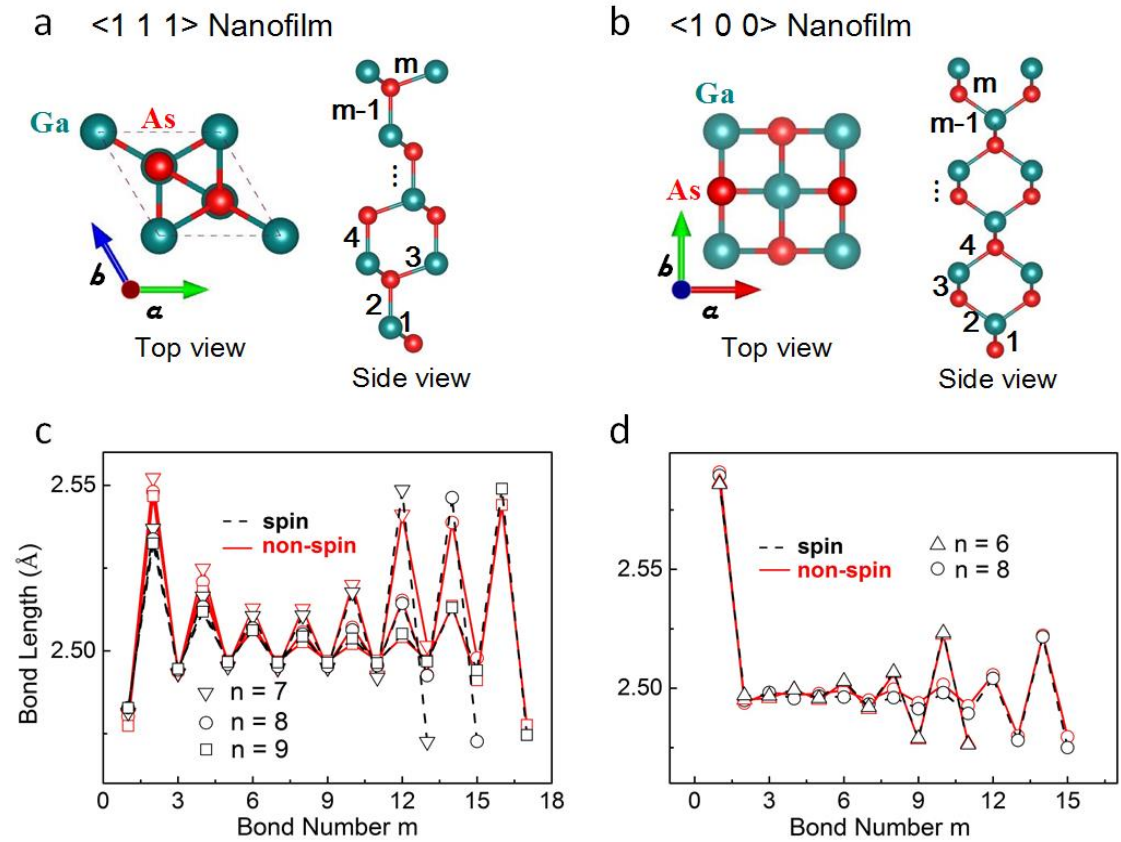

**FIG. 1.** Structural properties of GaAs nanofilms. The top view (left) and side view (right) of the single unit atomic structures of (a) <111> and (b) <100> nanofilms. **m** indexes bond number in the nanofilms. (c),(d) Bond length of <111> and <100> nanofilms along the <001> direction, respectively.

We calculate the ferromagnetic (FM) states of <111> nanofilms firstly, the results of which show clear evidences of surface magnetism as shown in Fig. 2(e). Based on this, the in-plane magnetism of two outmost surfaces is considered here since they contribute the vast majority of magnetic moments in the entire nanofilm. Following four possible in-plane magnetic ground states are considered in 2x2 surpercells: nonmagnetic (NM), ferromagnetic (FM)-α, antiferromagnetic (AFM)-α, and AFM-β, as shown in Fig. 2(a). It is shown that the FM-α configuration is the most favorable in-plane magnetic state in energy for all these surfaces. For Ga-surface of $Ga_5As_5$<111> nanofilm, the ground energy of FM-α configuration is 23.1 meV, 21.0 meV, and 20.9 meV per atom lower than that of the NM, AFM-α, and AFM-β, respectively. For As-surface of $Ga_5As_5$<111> nanofilm, the ground energy of FM-α configuration is 18.3 meV, 16.2 meV, and 16.2 meV per atom lower than that of the NM, AFM-α, and AFM-β, respectively. For <111> nanofilms thicker than 3 MLs, the energy difference between FM-α and NM states increases from 0 of 3 MLs to around 20 meV per atom of 5 MLs, and remains unchanged with further increasing **n**. The energy difference between FM-α and two AFM states shows the same trend. For <100> nanofilms, the ground states of Ga-surface and the closest As-layer are also found to be ferromagnetism, and the ground energy is around 25 meV, and 17 meV per atom lower than that of the NM, and two AFMs, respectively.

The exchange interaction between two different surfaces is considered as well. The two surfaces of <111> nanofilms have same spin direction as denoted by FM, or different spin directions as denoted by AFM in Fig. 2(b). The coupling between two surface states of the magnetic nanofilms thicker than 3 MLs is FM in the ground state. Results of the energy difference between the FM and AFM or NM states, and total magnetic moment (MM) are shown in Table I, where $\Delta E_{f-n} = E_{fm} - E_{nm}$ and $\Delta E_{f-a} = E_{fm} - E_{afm}$ per unit as shown in Figs. 1(a) and 1(b), respectively. $E_{fm}$, $E_{afm}$ and $E_{nm}$ represent the total energy of FM, AFM and NM configurations per unit, respectively. It is shown that both $\Delta E_{f-n}$ and $\Delta E_{f-a}$ have a sudden jump with increasing **n** from 3 to 5, and then become relatively stable. Meanwhile, the total magnetic moment shows the same trend as well, as shown in Table I. The detailed information about the thickness effect will be described below. These results were further confirmed in the larger 3×3 surpercells to eliminate the potential errors induced by small cell sizes. For the <100> case, the ferromagnetic coupling between two magnetic layers is much stronger than <111> nanofilms since they are located next to each other as shown in Fig. 2(f).

**TABLE I.** The energy difference per unit between the FM and NM states ($\Delta E_{f-n}$), the energy difference per unit between the FM and AFM states ($\Delta E_{f-a}$) and the total magnetic moments (MM) per unit of <111> and <100> nanofilms with various thickness.

|  | <1 1 1>$_3$ | <1 1 1>$_4$ | <1 1 1>$_5$ | <1 1 1>$_9$ | <1 1 1>$_{15}$ | <1 0 0>$_8$ |
|---|---|---|---|---|---|---|
| $\Delta E_{f-n}$ (meV) | 20.4 | -6.44 | -43.4 | -47.7 | -39.4 | -56.0 |
| $\Delta E_{f-a}$ (meV) | 20.2 | -1.17 | -7.8 | -9.5 | -6.8 | -20.1 |
| MM($\mu_B$/a.u.) | 0 | 0.544 | 0.896 | 1.342 | 1.381 | 0.438 |

It should be pointed out that different from the cubic structures showing semiconducting characters, nanofilms of GaAs are all ferromagnetic metals with flat bands crossing through the Fermi level ($E_F$). As the total density of states (DOS) of spin-polarized calculation shown in Figs. 2(c) and 2(d), both spin-polarization induced up- and down-spin electrons concentrate in the energy window of -1 eV and 1 eV and they are degenerated around the $E_F$. By visualizing the magnetization density in Figs. 2(e) and 2(f), we found the degenerated DOS in GaAs nanofilms originating from the strong localized states of surface atoms with almost entire unpaired electron wave function in *p* character. For $Ga_6As_6$<111> nanofilm, further analysis on the projected DOS of the outmost Ga and As atoms in Fig. 2(e) shows that spin-polarized electrons of the $4p_z$ orbital in outmost Ga atom presents a large splitting energy of 0.598 eV, and that of the $4p_z$ orbital in outmost As atom is 0.448 eV, rendering the imbalance redistribution of spin-polarized electrons at surfaces. The degenerated DOS induced by split $4p_z$ orbital in outmost Ga is also observed in the surface atoms of $Ga_6As_6$<100> nanofilm, and the closest As to Ga-surface has a large splitting energy of the $4p_x$ and $4p_y$ orbitals as well, with corresponding splitting energy of 0.161 eV for the $4p_z$ orbital of Ga and 0.160 eV, 0.165 eV for the $4p_x$ and $4p_y$ orbitals of As, respectively. Figure 2(f) shows the results of $Ga_6As_6$<100> nanofilm, and only $4p_x$ orbital of As is presented since $4p_y$ orbital has a similar DOS with $4p_x$, see Supplemental Material Fig. S2.

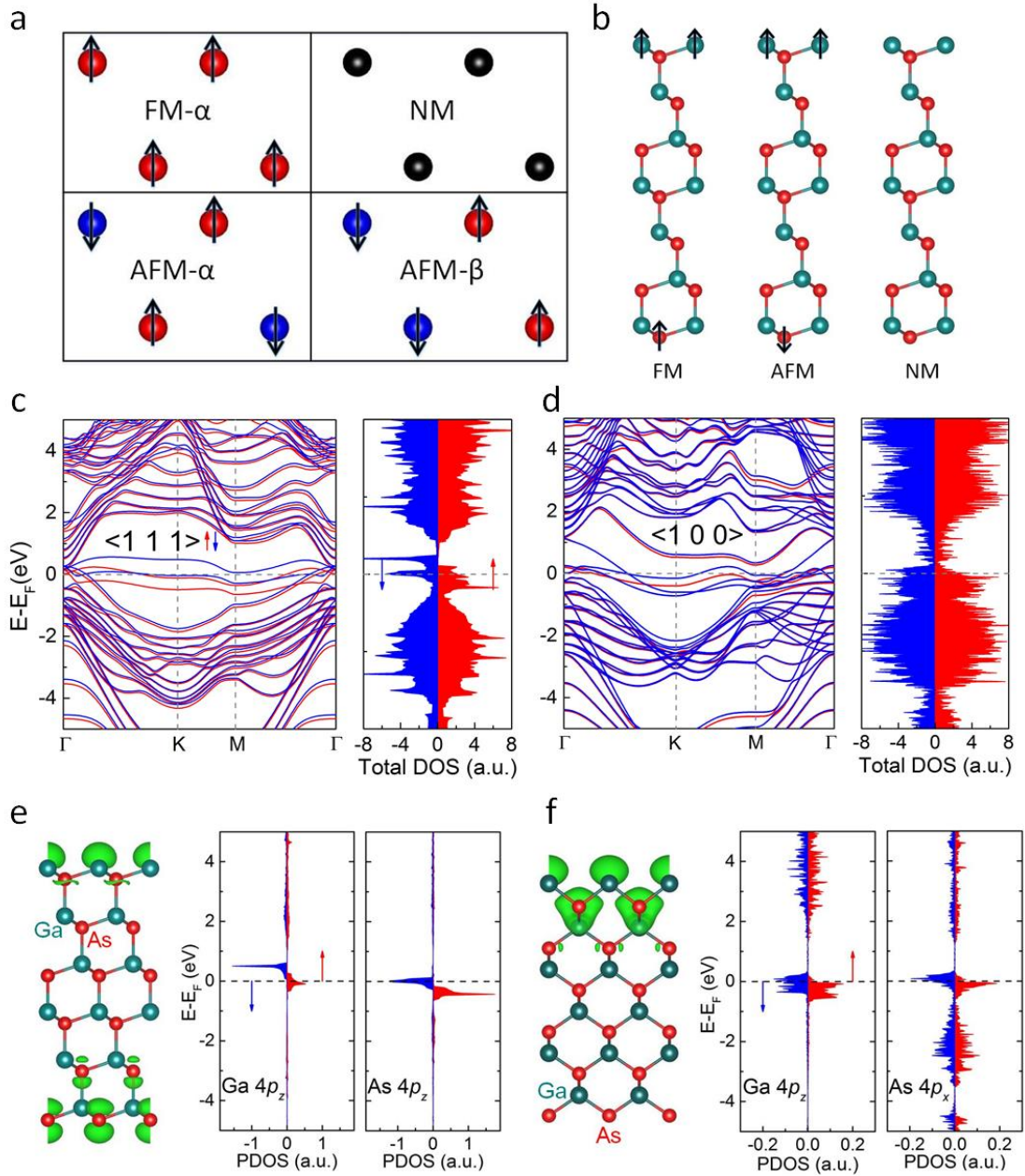

**FIG. 2.** Magnetic structure and Electronic properties of GaAs nanofilms. (a) Top view of the different in-plane spin configurations in <111> nanofilms: the FM-α ordered, NM ordered, AFM-α ordered, and AFM-β ordered, with only the surface atoms being shown. Spin directions are represented by arrows and colors (red for up, and blue for down). (b) Side view of the different inter-surfaces spin configurations in <111> nanofilms: the FM ordered, AFM ordered, NM ordered. Spin directions are represented by arrows. (c),(d) Spin-polarized band structure and total density of states in <111> and <100> nanofilms. The blue and red lines and arrows index the results for down- and up-spin, respectively. (e),(f) Spin-polarized electron distribution and projected density of states of the outmost Ga and As atoms of <111> nanofilm and <100> nanofilm, respectively.

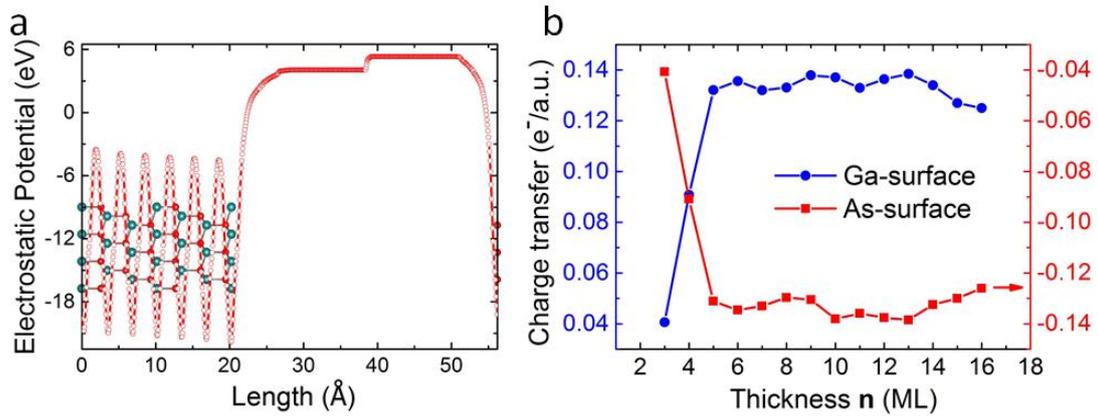

**FIG. 3.** Built-in electric field and charge transfer in the <111> nanofilms. (a) Plane-averaged electrostatic potential along the normal direction of $Ga_7As_7$<111> nanofilm. (b) The total charge-transfer between the surfaces as a function of thickness, represented by the sum of variations of valence electrons in four atoms located at the terminated and sub-terminated layers comparing to their bulk values in a single unit. The blue and red lines indicate the results of Ga-surface and As-surface, respectively.

To elucidate the origin of the large spin splitting, we plot the plane-averaged electrostatic potential along the normal of $Ga_7As_7$<111> nanofilm in Fig. 3(a). A linear distribution of the electrostatic potential is induced by electric-polarization, which means a built-in electric field pointing from the Ga-surface to the As-surface. The averaged potential difference between the outmost atoms at the Ga-surface and

As-surface is estimated to be 1.302 eV. The analysis on the plane-integrated charge transfer upon formation of the nanofilm from isolated atoms shows that electrons indeed deplete at the As-surface atoms and accumulate at the Ga-surface atoms. In Fig. 3(b) is the total charge transfer between two surfaces as a function of thickness, represented by the sum of variations of valence electrons in four atoms, which located at the terminated and sub-terminated layers, comparing to their bulk values. The valence electrons of As and Ga atoms in the bulk are 5.6 $e^-$ and 12.4 $e^-$, respectively. There are at least 0.12 $e^-$ transferring from As-surface to Ga-surface after a sudden jump with increasing **n** from 3 to 5 due to the built-in electric field. The sudden change is actually the formation process of spontaneous electric-polarization with the nanofilms getting thicker. An inspiring study also revealed thickness dependence of carrier density of the two-dimensional electron gas in SrTiO3 (111) slabs [79]. Usually, to satisfy the Stoner criterion, high density of states of p orbits at the Fermi level is difficult to realize. However, in <111> nanofilms, carriers at the Fermi level is dense enough at both As-surface and Ga-surface due to the charge transfer, which makes it an excellent candidate for realizing Stoner ferromagnetism. For <100> nanofilms, the magnetic moments mainly reside in the Ga-surface atoms and their origin is related to the existence of p electrons of well-defined spin polarization.

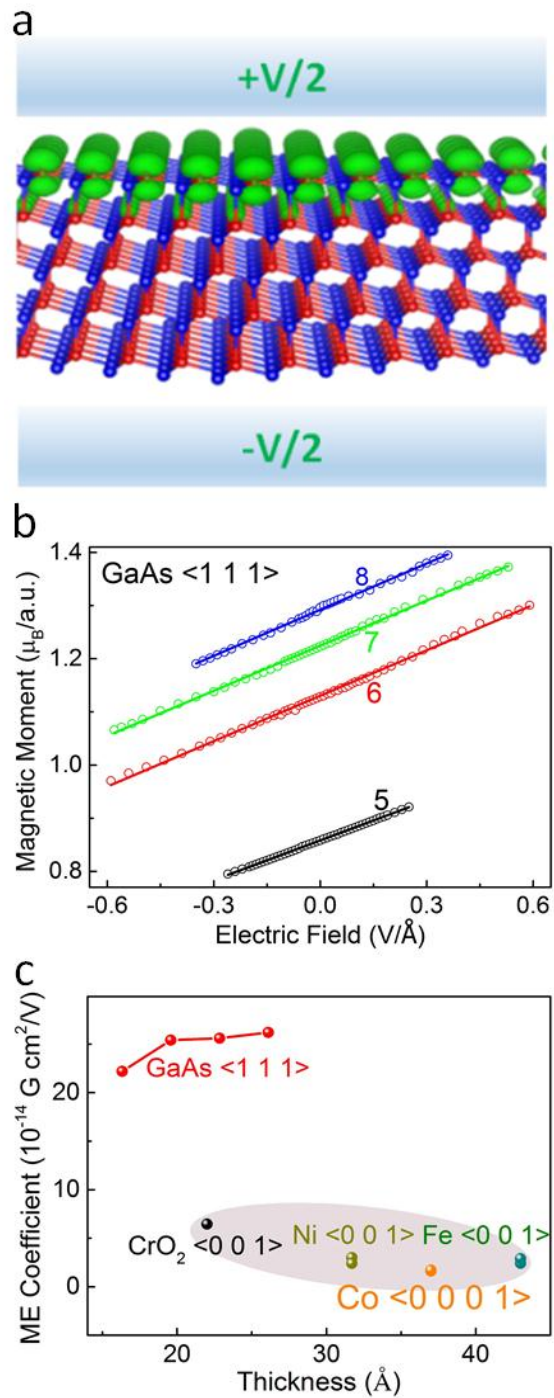

**FIG. 4.** ME effect in <111> nanofilms. (a) Schematic diagram of the nanofilms under vertically applied electric field. (b) External electric field induces linear magnetic polarization in <111> nanofilms. The circle with different colors represents nanofilms with various thickness, and the solid line indicates the fitted date of the magnetic moment in the nanofilms. (c) Comparison of the ME coefficient in this study and previous reported ferromagnetic metal and transition metal oxides nanofilms (Ref. [44,46]).

Considering that spin-polarized electrons at surfaces are driven by the electric-polarization across the nanofilms, we further apply the vertical electric field ($E_{ext}$) as shown in Fig. 4(a), which can modulate the electron transfer and tune the magnetic properties of the nanofilms. Here, we mainly focus on the ME effect in <111> nanofilms. Our results reveal perfect linear ME effects in <111> nanofilms as shown in Fig. 4(b). By fitting the calculated data, the obtained ME coefficient α in Eq. (1) is in the magnitude of $10^{-13}$ G·cm$^2$/V, which suggests the surface magnetism in <111> nanofilms is highly sensitive to externally applied electric field. When $E_{ext}$ increases from -0.6 to 0.6 V/Å, the magnetic moment of Ga$_6$As$_6$<111> increases from 0.960 to 1.310 $\mu_B$. For nanofilms with thickness varying from 5 to 8 MLs, the corresponding ME coefficient (in units of $10^{-13}$ G·cm$^2$/V) is estimated to be α = 2.22, 2.54, 2.56, and 2.57, respectively. The ME coefficient of <111> nanofilms is tens of times higher than those obtained at the ferromagnetic metal Co and Fe films [44], see Fig. 4(c). Thereby, <111> nanofilms are of greater potential in spintronic devices than traditional ferromagnetic metals. The high ME coefficient of <111> nanofilms is reasonable because it has a small plane-averaged electrostatic potential, which makes its surface magnetism much easier to tune. Since the surface magnetism is mainly accounted for the outmost surfaces atoms, the ME coefficient hardly changes with the thickness of <111> nanofilms increasing. As for <100> nanofilms, the surface magnetism also varies monotonously with the external electric field, see

Supplemental Material Fig. S3, but cannot show a linear relationship.

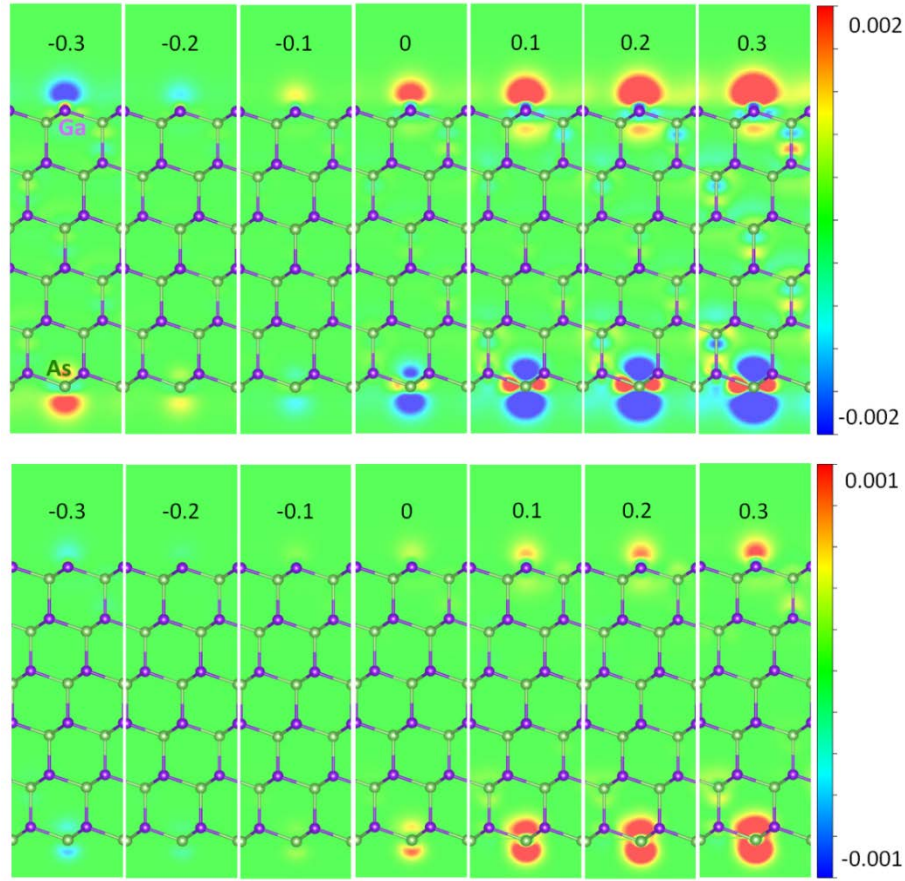

**FIG. 5.** Electric field modulation of transferring charge and spin density in $Ga_6As_6$<111> nanofilm. <110> slice of the differential charge (top panel) and spin-polarized electrons (bottom panel) under vertically applied electric field. Blue and red indicate electron depletion and accumulation, respectively.

The surface magnetism in <111> nanofilms is from the transferring charge and spin-polarized electrons at surfaces, the redistribution of which under the external electric field is contributed to the linear ME effect. Therefore, we present the differential charge $\Delta\rho$ of $Ga_6As_6$<111> nanofilm in Fig. 5 to show that, where the differential charge is defined as $\Delta\rho = \rho_e - \rho_0$, $\rho_e$ and $\rho_0$ are the charge density of $Ga_6As_6$<111> nanofilm with and without electric field, respectively. For $E_{ext}$ = 0 V/Å, redistribution of transferring charge originates from the spontaneous polarization across the nanofilm. When an external electric field is applied, the static equilibrium of transferring charge is broken. Under positive electric field ($E_{ext}$ is parallel to the built-in electric field), electrons will deplete (concentrate) at the outmost As-surface (Ga-surface) atoms, and the transferring electrons from As-surface to Ga-surface increase as the electric field strength increases. Oppositely, when negative electric field is applied ($E_{ext}$ is antiparallel to the built-in electric field), electrons transfer from the outmost Ga atoms to the outmost As atoms. Further analysis on the spin density difference reveals that those transferring electrons are not contributed equally to the up- ($n^\uparrow$) and down-spin ($n^\downarrow$) electrons. Under positive electric field, transferring electrons mainly occupy the *p* orbital as up-spin electrons and the occupied electrons increase with increasing electric field strength; under negative electric field, transferring electrons occupy the *p* orbital as down-spin electrons and the occupied electrons decrease as the electric field strength decreases. Since the magnetization density is defined as the difference between $n^\uparrow$ and $n^\downarrow$ as $M = n^\uparrow - n^\downarrow$, the total magnetic moment will monotonously increase with increasing electric field.

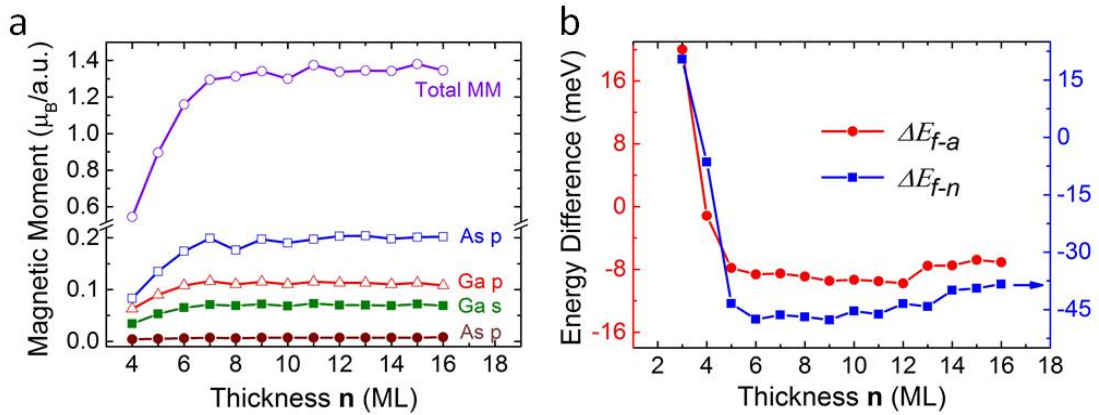

**FIG. 6.** Thickness-dependent of surface magnetism and relative stability in <111> nanofilms. (a) The thickness dependence of magnetic moments of <111> nanofilms and different orbitals in terminal atoms. (b) The energy difference between the nonmagnetic and ferromagnetic states in blue solid and the energy difference between the antiferromagnetic and ferromagnetic states in red circle.

As shown in Fig. 6(a), surface magnetism in GaAs nanofilms is thickness-dependent. Both total magnetic moment and projected magnetizations show constant value after a process of increasing. Meanwhile, the $\Delta E_{f-n}$ and $\Delta E_{f-a}$ have a sudden decrease with increasing **n** from 3 to 5, and then become relatively stable as shown in Fig. 6(b). The constant of $\Delta E_{f-n}$ and magnetic moment in thicker nanofilms is quite understandable because the saturation of transferring charge is reached, and the increasing or decreasing processes before that can be explained by the thickness dependence of charge transfer as shown in Fig. 3(b). As for the thickness dependence of $\Delta E_{f-a}$, long-ranged interlayer coupling between two surfaces should be considered. Recently, the importance of overlooked long-ranged interactions between magnetic ions has been revealed theoretically and experimentally, the objects of which contain both monolayers [80] and van der Waals crystals [81-83]. The <111> nanofilm of 3 MLs is an antiferromagnet with two surfaces having different spin directions due to significant contribution from interlayer superexchange interaction. For nanofilms thicker than 4 MLs, the surface states decouple and the $\Delta E_{f-a}$ may due to the contribution from the transferring charge between surfaces. A clear evidence is that applying every 0.2 V/Å positive electric field (parallel to built-in electric field) would leads to about 4 meV per unit decrease of $\Delta E_{f\text{-}a}$, indicating a more stable FM configuration. Due to the limited computation ability, the thickest nanofilm in our simulations is 5.0 nm ($Ga_{16}As_{16}$<111>) with the magnetic moment up to 1.35 $\mu_B$. The projected magnetization of outmost Ga and As atoms shown in the Fig. 6(a) further confirms that the $p$ orbital of terminal atoms is mainly contributed to the surface magnetism. The variation of the surface magnetism in <100> nanofilms shows the same trend but with a much smaller amplitude as shown in the supplemental material Fig. S5.

## IV. CONCLUSIONS

In summary, our first-principle calculations predict the existence of surface magnetism in metallic GaAs<111> and <100> nanofilms. The surface magnetism is attributed to the imbalance of spin-polarized electrons near the Fermi level. Built-in electric field induced by electric-polarization drives electrons aggregating at the outmost surface atoms, and it plays an important role in realizing and retaining the surface magnetism. Under an perpendicular electric field, the surface magnetization of these nanofilms significantly changes with the field strength, especially for <111> nanofilms exhibiting a strong linear magnetoelectric effect with high coefficients, which should be interesting for spintronic device development.

## ACKNOWLEGEMENTS

This work was supported by 973 program (2013CB932604), National Natural Science Foundation of China (51535005, 51472117), the Research Fund of State Key

Laboratory of Mechanics and Control of Mechanical Structures (MCMS-0416K01, MCMS-0416G01), the Fundamental Research Funds for the Central Universities (NP2017101), and a Project Funded by the Priority Academic Program Development of Jiangsu Higher Education Institutions.

---

See Supplemental Material at [URL will be inserted by publisher] for:
Structural properties of GaAs nanofilms; Projected density of states of the outmost As $4p_y$ orbit of <100> nanofilm; Magnetoelectric effect in GaAs<100> nanofilms; Thickness effect on the surface magnetism of GaAs<100> nanofilms.

[1] U. Gradmann, Journal of Magnetism and Magnetic Materials **6**, 173 (1977).

[2] S. Ohnishi, A. Freeman, and M. Weinert, Physical Review B **28**, 6741 (1983).

[3] M. Alde, S. Mirbt, H. L. Skriver, N. Rosengaard, and B. Johansson, Physical Review B **46**, 6303 (1992).

[4] M. Bender, D. Ehrlich, I. Yakovkin, F. Rohr, M. Baumer, H. Kuhlenbeck, H.-J. Freund, and V. Staemmler, Journal of Physics: Condensed Matter **7**, 5289 (1995).

[5] S. Apsel, J. Emmert, J. Deng, and L. Bloomfield, Physical review letters **76**, 1441 (1996).

[6] M. A. Ruderman and C. Kittel, Physical Review **96**, 99 (1954).

[7] T. Kasuya, Progress of theoretical physics **16**, 45 (1956).

[8] K. Yosida, Physical Review **106**, 893 (1957).

[9] A. Freeman and C. Fu, Journal of applied physics **61**, 3356 (1987).

[10] O. Eriksson, A. Boring, R. Albers, G. Fernando, and B. Cooper, Physical Review B **45**, 2868 (1992).

[11] I. M. Billas, A. Chatelain, and W. A. de Heer, Science **265**, 1682 (1994).

[12] Y.-W. Son, M. L. Cohen, and S. G. Louie, Nature **444**, 347 (2006).

[13] V. Barone and J. E. Peralta, Nano letters **8**, 2210 (2008).

[14] A. R. Botello-Méndez, F. López-Urías, M. Terrones, and H. Terrones, Nano letters **8**, 1562 (2008).

[15] Y. Li, Z. Zhou, S. Zhang, and Z. Chen, Journal of the American Chemical Society **130**, 16739 (2008).

[16] S. C. Erwin and F. J. Himpsel, arXiv preprint arXiv:1008.5358 (2010).

[17] J. Yu and W. Guo, The journal of physical chemistry letters **4**, 1856 (2013).

[18] G. Z. Magda, X. Jin, I. Hagymási, P. Vancsó, Z. Osváth, P. Nemes-Incze, C. Hwang, L. P. Biro, and L. Tapaszto, Nature **514**, 608 (2014).

[19] O. V. Yazyev and L. Helm, Physical Review B **75**, 125408 (2007).

[20] O. V. Yazyev, Physical review letters **101**, 037203 (2008).

[21] C. Jin, F. Lin, K. Suenaga, and S. Iijima, Physical review letters **102**, 195505 (2009).

[22] O. V. Yazyev and M. Katsnelson, Physical Review Letters **100**, 047209 (2008).

[23] D. Yu, E. M. Lupton, H. Gao, C. Zhang, and F. Liu, Nano Research **1**, 497 (2008).

[24] H. Fu, Z. Liu, C. Lian, J. Zhang, H. Li, J.-T. Sun, and S. Meng, Physical Review B **94**, 035427 (2016).

[25] I. Elfimov, S. Yunoki, and G. Sawatzky, Physical Review Letters **89**, 216403 (2002).

[26] S. Gallego, J. Beltrán, J. Cerdá, and M. Munoz, Journal of Physics: Condensed Matter **17**, L451 (2005).

[27] I. Shein and A. Ivanovskii, Physics Letters A **371**, 155 (2007).

[28] K. Janicka, J. P. Velev, and E. Y. Tsymbal, Journal of Applied Physics **103**, 378 (2008).

[29] H. Wu, A. Stroppa, S. Sakong, S. Picozzi, M. Scheffler, and P. Kratzer, Physical review

letters **105**, 267203 (2010).

[30] G. Fischer, N. Sanchez, W. Adeagbo, M. Lüders, Z. Szotek, W. M. Temmerman, A. Ernst, W. Hergert, and M. C. Muñoz, Physical Review B **84**, 205306 (2011).

[31] N. Sanchez, S. Gallego, and M. Munoz, Physical review letters **101**, 067206 (2008).

[32] H. Peng, H. Xiang, S.-H. Wei, S.-S. Li, J.-B. Xia, and J. Li, Physical review letters **102**, 017201 (2009).

[33] N. Sanchez, S. Gallego, J. Cerdá, and M. Muñoz, Physical Review B **81**, 115301 (2010).

[34] T. Cao, Z. Li, and S. G. Louie, Physical review letters **114**, 236602 (2015).

[35] P. Lunkenheimer *et al.*, arXiv preprint arXiv:1111.2752 (2011).

[36] S.-i. Ohkoshi, Y. Hamada, T. Matsuda, Y. Tsunobuchi, and H. Tokoro, Chemistry of Materials **20**, 3048 (2008).

[37] J. Coey, K. Wongsaprom, J. Alaria, and M. Venkatesan, Journal of Physics D: Applied Physics **41**, 134012 (2008).

[38] Y. Cai, Q. Ke, G. Zhang, and Y.-W. Zhang, The Journal of Physical Chemistry C **119**, 3102 (2015).

[39] J. Hoffman, I. Tung, B. Nelson-Cheeseman, M. Liu, J. Freeland, and A. Bhattacharya, Physical Review B **88**, 144411 (2013).

[40] N. A. Spaldin and M. Fiebig, Science **309**, 391 (2005).

[41] D. Astrov, Sov. Phys. JETP **13**, 729 (1961).

[42] M. Fiebig, Journal of Physics D: Applied Physics **38**, R123 (2005).

[43] K. Wang, J.-M. Liu, and Z. Ren, Advances in Physics **58**, 321 (2009).

[44] C.-G. Duan, J. P. Velev, R. F. Sabirianov, Z. Zhu, J. Chu, S. S. Jaswal, and E. Y. Tsymbal,

Physical review letters **101**, 137201 (2008).

[45] C.-G. Duan, C.-W. Nan, S. S. Jaswal, and E. Y. Tsymbal, Physical Review B **79**, 140403 (2009).

[46] Z. Zhang, C. Chen, and W. Guo, Physical review letters **103**, 187204 (2009).

[47] C.-G. Duan, S. S. Jaswal, and E. Y. Tsymbal, Physical Review Letters **97**, 047201 (2006).

[48] W. Eerenstein, M. Wiora, J. Prieto, J. Scott, and N. Mathur, Nature materials **6**, 348 (2007).

[49] S. Sahoo, S. Polisetty, C.-G. Duan, S. S. Jaswal, E. Y. Tsymbal, and C. Binek, Physical Review B **76**, 092108 (2007).

[50] F. Zavaliche, T. Zhao, H. Zheng, F. Straub, M. Cruz, P.-L. Yang, D. Hao, and R. Ramesh, Nano letters **7**, 1586 (2007).

[51] J. M. Rondinelli, M. Stengel, and N. A. Spaldin, Nature Nanotechnology **3**, 46 (2008).

[52] T. Kimura, T. Goto, H. Shintani, K. Ishizaka, T.-h. Arima, and Y. Tokura, nature **426**, 55 (2003).

[53] W. Eerenstein, N. Mathur, and J. F. Scott, nature **442**, 759 (2006).

[54] M. Bibes and A. Barthélémy, Nat. Mater **7**, 425 (2008).

[55] A. Ohtomo, D. Muller, J. Grazul, and H. Y. Hwang, nature **419**, 378 (2002).

[56] A. Ohtomo and H. Hwang, Nature **441**, 120 (2006).

[57] A. Brinkman *et al.*, arXiv preprint cond-mat/0703028 (2007).

[58] R. Oja *et al.*, Physical review letters **109**, 127207 (2012).

[59] Z. Zhang and W. Guo, Nano letters **12**, 3650 (2012).

[60] J. Blakemore, Journal of Applied Physics **53**, R123 (1982).

[61] H. Ohno, science **281**, 951 (1998).

[62] H. Zaari, M. Boujnah, H. Labrim, B. Khalil, A. Benyoussef, and A. El Kenz, Journal of superconductivity and novel magnetism **26**, 2961 (2013).

[63] H. Ohno, A. Shen, F. Matsukura, A. Oiwa, A. Endo, S. Katsumoto, and Y. Iye, Applied Physics Letters **69**, 363 (1996).

[64] T. Hayashi, M. Tanaka, T. Nishinaga, H. Shimada, H. Tsuchiya, and Y. Otuka, Journal of crystal growth **175**, 1063 (1997).

[65] S. D. Sarma, E. Hwang, and A. Kaminski, Physical Review B **67**, 155201 (2003).

[66] W. Liu *et al.*, Nature communications **7** (2016).

[67] M. R. Mahani, A. Pertsova, and C. M. Canali, Journal of Physics: Condensed Matter **26**, 394006 (2014).

[68] M. R. Mahani, M. F. Islam, A. Pertsova, and C. M. Canali, Physical Review B **89**, 165408 (2014).

[69] J. Wenisch *et al.*, Physical review letters **99**, 077201 (2007).

[70] G. Kresse and J. Furthmüller, Physical review B **54**, 11169 (1996).

[71] G. Kresse and J. Furthmüller, Computational Materials Science **6**, 15 (1996).

[72] D. Vanderbilt, Physical Review B **41**, 7892 (1990).

[73] H. J. Monkhorst and J. D. Pack, Physical review B **13**, 5188 (1976).

[74] J. Neugebauer and M. Scheffler, Physical Review B **46**, 16067 (1992).

[75] L. Bengtsson, Physical Review B **59**, 12301 (1999).

[76] H. L. Zhuang, A. K. Singh, and R. G. Hennig, Physical Review B **87**, 165415 (2013).

[77] C. L. Freeman, F. Claeyssens, N. L. Allan, and J. H. Harding, Physical review letters **96**,

066102 (2006).

[78] D. Wu, M. Lagally, and F. Liu, Physical review letters **107**, 236101 (2011).

[79] N. Sivadas, H. Dixit, V. R. Cooper, and D. Xiao, Physical Review B **89**, 075303 (2014).

[80] N. Sivadas, M. W. Daniels, R. H. Swendsen, S. Okamoto, and D. Xiao, Physical Review B **91**, 235425 (2015).

[81] B. Huang *et al.*, Nature **546**, 270 (2017).

[82] M. A. McGuire, H. Dixit, V. R. Cooper, and B. C. Sales, Chemistry of Materials **27**, 612 (2015).

[83] W.-B. Zhang, Q. Qu, P. Zhu, and C.-H. Lam, Journal of Materials Chemistry C **3**, 12457 (2015).